\documentclass{river-journal}
\usepackage{rivps}

\usepackage{comment}
\usepackage[hidelinks]{hyperref}
\usepackage{xurl}
\usepackage[numbers]{natbib}

\usepackage[T1]{fontenc}

\usepackage{xcolor}

\definecolor{editorGray}{rgb}{0.95, 0.95, 0.95}
\definecolor{editorOcher}{rgb}{1, 0.5, 0} 
\definecolor{editorGreen}{rgb}{0, 0.5, 0} 
\usepackage{upquote}

\usepackage{listings}

\lstdefinelanguage{JavaScript}{
  morekeywords={typeof, new, true, false, catch, function, return, null, catch, switch, var, if, in, while, do, else, case, break},
  morecomment=[s]{/*}{*/},
  morecomment=[l]//,
  morestring=[b]",
  morestring=[b]'
}

\lstdefinelanguage{HTML5}{
        language=html,
        sensitive=true, 
        alsoletter={<>=-},
        otherkeywords={
        <html>, <head>, <title>, </title>, <meta, />, </head>, <body>,
        <canvas, \/canvas>, <script>, </script>, </body>, </html>, <!, html>, <style>, </style>, ><
        },  
        ndkeywords={
        =,
        charset=, id=, width=, height=,
        border:, transform:, -moz-transform:, transition-duration:, transition-property:, transition-timing-function:
        },  
        morecomment=[s]{<!--}{-->},
        tag=[s]
}

\usepackage{tabularx}

\usepackage[capitalise]{cleveref}

\newcolumntype{z}{>{\hsize=.25\hsize}X}
\newcolumntype{s}{>{\hsize=.5\hsize}X}

\usepackage{pgfplots}
\pgfplotsset{compat=1.18}
\usepgfplotslibrary{statistics} 

\usepgfplotslibrary{colorbrewer}
\pgfplotsset{compat = 1.15, 
             cycle list/Dark2} 
\usetikzlibrary{pgfplots.statistics, pgfplots.colorbrewer}

\usepackage{graphicx}
\graphicspath{{./figures/}}

\usepackage{booktabs, multirow} 
\usepackage{soul}
\usepackage{xcolor,colortbl} 
\usepackage{changepage,threeparttable} 
\usepackage{adjustbox}

\usepackage[english]{babel}
\addto\extrasenglish{

}

\makeatletter
\newcommand*{\centerfloat}{%
  \parindent \z@
  \leftskip \z@ \@plus 1fil \@minus \textwidth
  \rightskip\leftskip
  \parfillskip \z@skip}
\makeatother

\usepackage{enumitem}
\setlist[description]{style=nextline}

\begin{document}

\begin{opening}
\title{The Case for HTML First Web Development}
\author{Juho Vepsäläinen}
\institute{Aalto University, Department of Computer Science, juho.vepsalainen@aalto.fi}
\end{opening}

\runningtitle{The Case for HTML First Web Development}
\runningauthor{J. Vepsäläinen}

\subsection*{Abstract}

Since its introduction in the early 90s, the web has become the largest application platform available globally.
HyperText Markup Language (HTML) has been an essential part of the web since the beginning, as it allows defining webpages in a tree-like manner, including semantics and content.
Although the web was never meant to be an application platform, it evolved as such, especially since the early 2000s, as web application frameworks became available. 
While the emergence of frameworks made it easier than ever to develop complex applications, it also put HTML on the back burner.
As web standards caught up, especially with milestones such as HTML5, the gap between the web platform and frameworks was reduced.
HTML First development emphasizes this shift and puts focus on literally using HTML first when possible, while encouraging minimalism familiar from the early days of the web.
It seems HTML-oriented web development can provide clear benefits to developers, especially when it is combined with complementary approaches, such as embracing hypermedia and moving a large part of application logic to the server side.
In the context of the htmx project, it was observed that moving towards HTML can reduce the size of a codebase greatly while leading to maintenance and development benefits due to the increased conceptual simplicity.
Holotype-based comparisons for content-oriented websites show performance benefits, and the same observation was confirmed by a small case study where the Yle website was converted to follow HTML First principles. 
In short, the HTML First approach seems to have clear advantages for web developers, while there are open questions related to the magnitude of the benefits and the alignment with the recent trend of AI-driven web development.


\vspace{6pt}
\setlength{\fboxsep}{6pt}
\fbox{%
  \parbox{0.9\columnwidth}{%
    This work has been submitted to Journal of Web Engineering (JWE) for possible publication. Copyright may be transferred without notice, after which this version may no longer be accessible.
  }%
}
\vspace{6pt}

\section{Introduction}
\label{sec:introduction}
The web, described by \citet{berners1994world} in 1994, encapsulated the idea of a global information exchange, and it more than succeeded in this vision.
Based on \citet{worldbankWorld2025}, nearly 70\% of the global population has access to the web today, making it the largest available application platform.
Although the web was initially meant specifically for websites sharing information, it morphed over time to be able to host web applications with interactive functionality to enable, for example, the emergence of social media.

\subsection{HTML - a key part of modern web development}

Typically, web applications consist of code written in three languages: CSS, HTML, and JavaScript.
CSS (Cascading Style Sheets) captures the concern of styling, or more specifically, visual outlook.
HTML, or HyperText Markup Language, defines a page structure using an XMLish\footnote{Unlike XML, HTML does not have as rigid a schema, although this was attempted earlier with XHTML, which went out of favor.} syntax, allowing users to define tree structures with data and semantics for each element of the tree.
JavaScript captures specifically the aspect of handling application logic in coordination with the Document Object Model (DOM).

HTML, illustrated in \autoref{fig:html}, is a quite expressive language, especially in its newest editions, and it allows capturing semantics valuable for defining complex user interfaces while taking care of common accessibility concerns.
Simultaneously, it is easy to overlook the possibilities enabled by HTML and lose out on its benefits.
So-called HTML First Manifesto \cite{htmlfirst2023} raised this point in 2023, and the observation that HTML can do far more than you might think as a developer is at the core of this paper.

\begin{figure}[h]
\centering
\begin{lstlisting}{html}
<section>
  <p>This is a paragraph of text</p>
  <img href="./demo.png" title="Demo image" />
</section>
\end{lstlisting}
\caption{Example of HTML markup with HTML elements, nesting, and attributes within an \texttt{img} element.}
\label{fig:html}
\end{figure}


\subsection{Related work}

As pointed out by \citet{saarinen2024understanding}, the evolution of the web was not without a cost since the current web has major opportunities for data savings estimated to be over half of the current amount of data utilized.
It can be argued that there may be several possible directions to address the issue, as highlighted by \citet{vepsalainen2023overview}.
In this paper, I investigate one potential direction, the possibility of leveraging the platform more effectively through so-called HTML First development.

\subsection{Research questions}

To study how the usage of HTML has evolved and how ideas like the HTML First Manifesto can shape it, I arrived at the following two research questions:

\begin{enumerate}
    \item What were the factors leading to the underuse of HTML features in web application development?
    \item How can HTML First development remedy the underuse of HTML, and why is it useful?
\end{enumerate}

The first question gives background to the discussion and helps to understand how we arrived at the present state.
The second question looks deeper into HTML First development and considers the implications for web development.

\subsection{Structure of the paper}

To address the first research question, I cover the path of HTML from a language meant for defining websites to one that allows defining web applications to a certain extent, while showing how web frameworks changed the landscape and how modern HTML can make the language relevant again in \autoref{sec:html-to-frameworks}.
The second research question is covered by the following two sections.
In \autoref{sec:html-patterns}, I cover common user interface patterns that are available in plain HTML and that can occasionally replace a significant amount of code.
In \autoref{sec:comparison}, I show how following HTML First principles can potentially improve websites on factors such as maintainability and performance. 
In \autoref{sec:discussion}, I consider the current state of HTML from different angles while discussing the findings.
I recap the questions and my answers to them in conclusion within \autoref{sec:conclusion}.

\section{From HTML to Web Frameworks and back}
\label{sec:html-to-frameworks}
HTML was an integral part of the definition of world wide web (www)\footnote{Since the web has become so widely used, it is common to refer to it as www or simply as the web, both of which can be used interchangeably.} by \citet{berners1994world} already in 1994.
HTML was derived from SGML (Standard Generalized Markup Language), an earlier language, as described by \citet{tabares2021}.
The point was to have a language independent of the browser, leaving interpretation to the user\footnote{CSS, defined later, addressed this concern and enabled consistent rendering across browsers over time. Interestingly enough, this was not a consideration in the original specification of www as styling was considered a user-level concern.}.
In this section, I will go through the main steps leading to modern HTML before considering how HTML First development fits the picture.

\subsection{From HTML to HTML5}

As the web grew in popularity and gained more functionality, the second era of the web, known as Web 2.0 -- the era of web applications -- began in the early 2000s \cite{tabares2021}.
Due to the relative maturity of the web platform, developers were able to program complex User Interfaces (UIs) on top of the standards to go beyond their limitations, and this, in turn, allowed the web to become accessible by an increasing number of users with the emergence of services driven by user-generated content \cite{tabares2021}.
While web standards lagged, proprietary and closed alternatives for handling multimedia content emerged, challenging the web's open nature and hindering standardization, creating a strong incentive to modernize HTML to capture the latest developments and avoid fragmentation of the web \cite{tabares2021}.
This development culminated in the standard HTML5, although arriving at HTML5 was not easy, as the W3C\footnote{W3C was founded as early as 1994 to coordinate the development of the web \cite{w3History}.} (World Wide Web Consortium) faced challenges.
W3C had different ideas about what should happen to HTML than major companies in the space, which coordinated themselves as the WHATWG (Web Hypertext Application Technology Working Group), leading to a standards battle \cite{tabares2021}.
Essentially, W3C was pushing for XHTML2, a version of HTML that would not have been backward-compatible and would have required abandoning earlier work \cite{tabares2021}.
At the same time, WHATWG advocated maintaining backwards compatibility and expanding HTML's capabilities to meet the demands of the web's application era \cite{tabares2021}.
As work on XHTML2 stalled after two and a half years, the W3C joined forces with WHATWG to finalize HTML5, which reached the official recommendation stage in 2014 and was later updated in 2017 \cite{tabares2021}.
Since 2019, HTML has been developed as a living standard by WHATWG in collaboration with W3C \cite{laakso2025}.
Instead of following a specific schedule, a living standard is updated continuously while copies of the standard are stored\footnote{HTML standard is available through GitHub at \url{https://github.com/whatwg/html}.} \cite{laakso2025}.

\subsection{The era of web applications (2000)}


Web 2.0, formalized in the early 2000s, marked a shift from websites to web applications \cite{o2009web}.
Technically, the era was characterized by the introduction of new techniques, such as Single Page Applications (SPAs), that enabled developers to build more complex applications \cite{vepsalainen2025emergence}.
To facilitate the easy creation of SPAs, several web frameworks emerged to address related concerns and work around the deficiencies of the web platform at the time.
Although frameworks enabled the development of complex web applications, they did not enforce best practices.
As a result, new problems emerged, such as the so-called div soup\footnote{In div soup, the semantics of the markup are lost at the cost of reduced accessibility and performance \cite{pfeiffer2014improving}.} \cite{pfeiffer2014improving}, emerged.
In other words, it was easy to forget the platform below and produce code that worked but was not ideal from an implementation standpoint.
HTML First manifesto addresses this problem by putting attention back to basics.

\subsection{The case for HTML First development (2023)}

Tony Ennis describes the HTML First manifesto \cite{htmlfirst2023} as a set of principles that highlight how modern HTML can replace earlier technical solutions designed to address deficiencies in the HTML specification.
The HTML First approach could be described as an antithesis to modern development practices, though it can be considered complementary to them, since HTML cannot meet everything a web application requires.
The point of the manifesto is to acknowledge the current state of HTML and specifically encourage developers to leverage modern HTML wherever possible.
HTML First manifesto \cite{htmlfirst2023} includes several principles I have covered in detail below:

\begin{enumerate}
    \item \textbf{Apply the principle of least power} - The idea of this principle is to favor technologies from simple to complex, starting from HTML markup through CSS rules to JavaScript code. In other words, the principle gives preference to simpler technologies while acknowledging that more complex ones may still be used as needed. This aligns with \citet{stewart2025}'s idea of progressive complexity.
    \item \textbf{Prefer "vanilla" approaches over external frameworks} - Many simple tasks, even with dynamic functionality\footnote{A classic example is a "show more" kind of button that opens further information when pressed.}, can be implemented with plain HTML, and therefore should be implemented using it.
    \item \textbf{Avoid build steps where possible} - As a part of the development of modern web tooling, a build step was introduced to the process. The idea of this principle is to avoid build steps as possible to simplify development.
    \item \textbf{Minimise unnecessary client-side state} - This principle goes against complex state in client-side enabled by SPAs and encourages developers to minimize client state to simplify their applications, for example, by persisting changes and performing validations at the server.
    \item \textbf{Retain the View-Source affordance} - Given that SPAs obscure their compilation steps and what happens at the browser, this principle encourages developers to work closer to web standards to enable easier debuggability directly at the browser.
    \item \textbf{If you must use libraries, prefer ones that re-purpose existing concepts over those that create their own lexicon} - This principle encourages developers to use simple libraries over complex frameworks to complement HTML where needed. A good example is simple state management libraries that extend HTML attribute semantics to make it more powerful.
\end{enumerate}

HTML First development is consistent with the idea of progressive enhancement introduced by \citet{gustafson2015adaptive} in 2003.
As \citet{wells2007progressive} describe, in progressive enhancement, the focus is put on page content, which is complemented by presentation, and finally client-side scripting.
This is roughly equivalent to the second principle of HTML First development and highlights the fact that similar concerns have existed earlier on the web.
Even before progressive enhancement, developers implemented a technique called graceful degradation, as early browsers lacked features and did not render the same markup consistently, making it necessary to accept this fact, especially in the context of early websites \cite{gustafson2015adaptive}.

\subsection{Progressive Complexity Manifesto (2025)}

In 2025, \citet{stewart2025} published a manifesto calling for progressive web complexity and the dissolution of the false dichotomy between websites and web applications, acknowledging that there is a vast middle ground in between.
The manifesto aligns with the HTML First manifesto.
However, it has a slightly different focus, asking developers to consider how much complexity they need in each web development use case.
This aligns well with the observation of \citet{vepsalainen2025emergence} that hybrid rendering models have their place in web development as they allow capturing this middleground.

\subsection{Summary}

HTML is one of the core technologies of the modern web, available since the beginning.
The history of HTML standardization is not straightforward, as it took some struggle to reach the current model of living standardization, where the HTML standard is updated without a regular schedule.
Web 2.0, with its web applications, placed reduced emphasis on HTML.
To counter this, the HTML First manifesto reminds us of the power of the platform, as the standards have caught up and can be used for more purposes than developers may even be aware of.
Similar sentiments have existed before in the form of progressive enhancement, but HTML First development formalizes the idea further as a modern minimalism that complements modern web application development practices.

\section{Common HTML Patterns}
\label{sec:html-patterns}
Describing page structure and its semantics is at the core of HTML.
However, beyond content-focused websites, HTML can cover a limited set of interactive functionality thanks to improvements made to the HTML standard that acknowledge the role of the web as an application platform.
As \citet{petros2023} shows, HTML is amenable to extension and a natural place for developers to extend its semantics through directives through custom attributes attached to HTML elements.
I will cover several examples of how to use HTML for common application-related tasks in this section, and although the list is not exhaustive, it should give you some idea of why investigating HTML First alternatives during development may be a worthwhile option.

\subsection{Foldable containers}

As discussed by \citet{mozilladetails2025}, \texttt{details} and \texttt{summary} HTML elements work in tandem to provide foldable disclosure containers.
Typically, these elements are rendered with their content hidden by default and with only the summary visible unless the \texttt{open} attribute has been specified for the \texttt{details} element.
The example at \autoref{fig:summary} based on \cite{mozilladetails2025} illustrates the usage.

\begin{figure}[h]
\begin{lstlisting}{html}
<details>
  <summary>What is HTML First?</summary>
  A paradigm that favors the usage of HTML before other
  technologies.
</details>
\end{lstlisting}
    \caption{\texttt{details} and \texttt{summary} elements work in tandem to enable foldable containers using pure HTML.}
    \label{fig:summary}
\end{figure}

\subsection{Semantic elements}

As illustrated by \citet{pfeiffer2014improving}, especially before HTML5 was introduced, so-called "div soup" was a common problem.
HTML5 specification \cite{hickson2011html5} improved the situation by including a group of semantic elements, for example \texttt{article}, \texttt{section}, \texttt{header}, \texttt{footer}, \texttt{nav}, and \texttt{aside}.
It can be argued that even today, many of these elements are underused, and this is one of the points that the HTML First approach emphasizes, as being aware of available elements and using them correctly is valuable.

\subsection{Form validation}

As shown by the \citet{whatwgHTMLStandard}, modern HTML provides versatile ways to validate forms on the client without JavaScript.
One interesting, and perhaps underused feature, is field autocompletion \cite{whatwgHTMLStandard} available through the \texttt{autocomplete} attribute that describes the purpose of the field and hints to the browser what kind of completion to use\footnote{This is particularly useful with repetitive information, such as name, address, etc. details as often browsers save this information for the user to reuse.}.
Modern form validation is widely supported \cite{caniuseFormValidation}, and it is one of those features that goes well with HTML First development.

Even though HTML allows validation, developers must handle validation at the server side to ensure application security.
Therefore, HTML form validation should be considered as a helpful feature but not a complete solution.
Ideally, validation rules should exist at the backend, and these rules could then be used to generate frontend validation helpers while still performing further validation of submission at the server logic.

\subsection{Dialogs and popovers}

Dialogs and popovers are commonly used UI features that are often used by web applications, and web standards have caught up in this regard by providing many types of APIs \cite{caniusequotdialogquotUse} to ease the creation of these UI elements easily.
Earlier developers had to resort to JavaScript for these features, but nowadays it may not be necessary at all, depending on which browsers you target.
Based on \cite{caniusequotdialogquotUse}, the main caveat seems to be that some of the dialog-related features are not available in mainstream Safari browsers at the time of writing.
However, this problem may have been fixed by the time you read this article.

\subsection{Implementing custom attributes}

As \citet{petros2023} discusses in his blog post, HTML can be extended in user space by attribute-based directives that extend its semantics as shown by \autoref{fig:directive}

\begin{figure}[ht]
\begin{lstlisting}{html}
<button alert message="I was clicked">
  Click me
</button>
<script>
// Get all the buttons with the 'alert' attribute
const buttons = document.querySelectorAll(
  'button[alert]'
);
buttons.forEach(button => {
  // Set the button to alert with the attribute
  // contents on click
  button.addEventListener('click', () => {
    alert(button.getAttribute('message') };
  );
})
</script>
\end{lstlisting}
    \caption{The example adapted from \citet{petros2023} shows how to implement reusable directives to extend HTML semantics.}
    \label{fig:directive}
\end{figure}

This extension could be extracted to a JavaScript file of its own and used across the whole codebase.
Although this way may feel counterintuitive for those used to alternative models, it can be argued that connecting logic directly with HTML leads to declarative code with higher cohesion, therefore improving maintainability, as there is only a single clear place to look, even if the code for specific directives has been extracted elsewhere.
It is this idea of extending HTML through attributes that led to the creation of light libraries around HTML that extend its capabilities in ways that allow developers to model application logic largely using HTML syntax.

\subsection{Complementing HTML with light libraries}

As the HTML First manifesto acknowledges that not everything can be implemented using pure HTML, the manifesto encourages the use of libraries for complementary purposes.
Although many libraries exist, I will show you two light, runtime-based ways to handle state using a library, namely Alpine.js and htmx\footnote{I have listed several more options at \url{https://sidewind.js.org/\#related-approaches}.}.
The core idea of Alpine.js is to capture state management on top of HTML itself using a set of attributes as seen in \autoref{fig:alpine}.
Conversely, htmx has a strict focus on communicating with servers and expanding the scope of HTML to allow hypermedia-style\footnote{You can consider hypermedia as a generalization of hypertext in the sense that it supports formats beyond text as discussed by \citet{tolhurst1995hypertext}.} programming through HTML attributes, as \autoref{fig:htmx} illustrates.

\begin{figure}[h]
\centering
\begin{minipage}{0.45\textwidth}
\begin{lstlisting}{html}
<div x-data="{open: false}">
  <button
    @click="open = true"
  >
    Expand
  </button>
  <span x-show="open">
    Content
  </span>
</div>
\end{lstlisting}
\caption{Note how \texttt{x-data}, \texttt{x-show}, and \texttt{@click} are used in tandem to manage state. The example has been adapted from \citet{alpinejs}.}
\label{fig:alpine}
\end{minipage}\hfill
\begin{minipage}{0.45\textwidth}
\begin{lstlisting}{html}
<button
  hx-post="/clicked"
  hx-swap="outerHTML"
>
  Click Me
</button>
\end{lstlisting}
\caption{When the button is clicked, htmx sends a \texttt{POST /clicked} request to the server and replaces itself with the result. The example has been adapted from \citet{htmx}.}
\label{fig:htmx}
\end{minipage}
\end{figure}

So-called Triptych proposals by \citet{petros2025} aim to integrate features like partial page replacement directly to HTML and, most importantly, make it more expressive for handling network requests beyond its current capabilities.
The argument is that by extending the semantics, HTML becomes able to handle REST \cite{fielding2000architectural}\footnote{REST is a common pattern for modeling web server APIs and \citet{swaggerOpenAPISpecification} captures its principles as a schema while Fielding's groundbreaking thesis \cite{fielding2000architectural} is the origin of the pattern.} completely expanding HTML's usefulness to cover a larger variety of problem spaces without the use of external libraries like htmx.

Both of the shown approaches show potential directions to expand HTML semantics using light libraries (both are in the 10-20 kB range when minified and gzipped).
Although they come with some learning and loading-related costs, these approaches are far lighter than leveraging full frameworks, and most importantly, they sit well on top of regular HTML.
It can be argued that HTML like this is not standard anymore, but libraries like this allow experimenting with what HTML could be like before standardizing on a specific direction.

\subsection{Summary}

The examples discussed in this section are far from exhaustive, and more libraries like these exist.
The key point is that many common UI-related problems can be solved with HTML already.
The situation keeps improving year by year as standardization is catching up and modeling missing APIs at the platform level, decreasing the need to reach out for JavaScript-based solutions.
It is likely inevitable that most complex UIs will require some amount of code, but even then, there is value in replacing common patterns with HTML and a bit of CSS where possible.

\section{Comparison}
\label{sec:comparison}
The HTML First approach has its benefits, particularly for content-oriented websites.
To illustrate the benefits from multiple angles, I gathered experiences from related case studies based on existing gray literature and performed comparisons of my own within web application holotypes\footnote{Holotypes are a concept from zoology that translates to classifying web applications \cite{vepsalainen2025potential}} before benchmarking a site I partially rewrote using HTML First principles\footnote{The code for my experiments is available at \url{https://github.com/bebraw/html-first-demos}.}.
These methods were chosen to study HTML-oriented web development from multiple angles as they allowed leveraging existing data while keeping the research grounded in reality through measurements.
The case studies are based on experiences of the htmx project that encourages the use of HTML First practices.
I chose the comparisons to show how existing sites can differ based on their orientation to either HTML or JavaScript.
The benchmark shows how simple modifications to an existing site can improve the site on multiple metrics.

\subsection{Prior case studies by htmx}

As discussed earlier in \autoref{sec:html-patterns}, htmx is a server-oriented library that expands the semantics of HTML to allow hypermedia-style development.
Developers of htmx have gathered several case studies showing the benefits of this HTML-oriented approach, and I have captured key insights from their collection below.

\begin{table}[h]
    \centering
    \caption{Case studies based on htmx project}
    \label{tab:placeholder_label}
    \begin{tabular}{p{3.0cm}p{6.5cm}c}
        \toprule
        Case & Experiences & Source \\
        \midrule
        Porting from React to htmx in a real-world Software as a Service (SaaS) product. & The refactoring required two months of effort. The size of the codebase was reduced by 67\% while the amount of Python code grew from 500 lines to 1200 lines. The amount of JavaScript dependencies was reduced by 96\%. The build time of the web portion was reduced by 88\% from 40 seconds to 5 seconds. At the client, the first load time-to-interactive was reduced to half while allowing the application to handle much larger datasets. On top, the memory usage of the application was reduced to roughly half of before. & \cite{gross2022} \\
        Porting from React to htmx in a real-world SaaS. & The size of the codebase was reduced by 61\% while the number of files was reduced by 72\% and the total number of file types by 38\%. The team was able to drop Figma from their workflow and develop their interfaces directly in HTML after the refactoring was completed. & \cite{gross2023} \\
        Porting from Next.js to htmx in an open-source URL shortener service. & The amount of project dependencies was reduced 87\% (24 to 3), the size of the codebase was reduced by 17\%, after refactoring, the project did not have a build step anymore, and the size of the website was reduced by more than 85\%. & \cite{ezzati2024} \\
        Porting from WebAssembly to htmx in a real-world SaaS product. & The refactoring required three weeks of intense work. The size of the codebase was reduced by 78\%, the number of Rust packages in the codebase was reduced from eight to one, and after the refactoring, the number of bug reports per week reduced from about five to one. In addition, it was reported that feature development was greatly accelerated after the changes. & \cite{fioti2025} \\
        \bottomrule
    \end{tabular}
\end{table}

The examples highlight how moving logic from the frontend to the backend can reduce the amount of code while often improving user experience.
It can be argued that now the logic exists in some other language than JavaScript, but even that can be considered a benefit, depending on the team.
That said, it is entirely possible that these positive cases were particularly good fits for relying more on HTML and leveraging a hypermedia-based approach.

There is also a documented case where it was clear that using htmx was not a good fit.
htmx was considered for Gumroad, a popular SaaS product, but the development team opted against it after understanding that the solution would be too limited given their high demands for interactive functionality and support provided by the ecosystem to avoid reinventing the wheel, amongst other factors \cite{lavingia2024}.
The experience of Gumroad highlights the fact that established approaches still have their place and refactoring may not always be sensible, especially if the possible benefits are not that clear.

\subsection{Comparisons within web application holotypes}

Web application holotypes, as described by \citet{vepsalainen2025potential}, provide a way to categorize websites based on their dynamic requirements and other features.
Common holotypes include portfolio sites, content sites, storefronts (e-commerce), social networks, and immersive applications \cite{vepsalainen2025potential}.
Most importantly, holotypes allow us to consider sites within a category relative to one another.

In this portion of the benchmark, I formed several site pairings within multiple categories to compare HTML-oriented and JavaScript-heavy websites\footnote{Note that since the benefits of the HTML First approach are more apparent in sites relying on content, the selection is biased towards content sites of different types. Highly interactive web applications often require a considerable amount of JavaScript code that cannot be avoided by definition.}.
This dichotomy is visible in the technical choices of the sites and especially what kind of files they require to work\footnote{Naturally, JavaScript-heavy sites may load a considerable amount of JavaScript while HTML-oriented sites may not even require it at all.}
It can be argued that the HTML-oriented sites could be optimized further based on HTML First principles.
Still, even in their current state, they give a good baseline for comparisons due to their focus on leveraging HTML over JavaScript.
I have listed the pairings in \autoref{tab:holotype_table}, with the first of each being considered as the HTML-oriented variant and the second one being the JavaScript-heavy one.

\begin{table}[ht]
    \caption{The table enumerates holotype pairs under consideration to evaluate the HTML First approach against the JavaScript-heavy one.}
    \label{tab:holotype_table}
    \begin{tabular}{cp{0.8cm}lp{2.8cm}}
        \toprule
        Holotype & Site & Link & Description \\
        \midrule
        Content & Learn Rust & \url{https://rust-lang.org/learn/} & Rust learning resource \\
        Content & Next.js & \url{https://nextjs.org/docs/} & Next.js documentation \\
        News & Hacker News & \url{https://news.ycombinator.com/} & Y Combinator technology news \\
        News & The Verge & \url{https://www.theverge.com/} & A popular tech news aggregator \\
        E-commerce & Kirby CMS & \url{https://getkirby.com/} & A content management system \\
        E-commerce & Shopify platform & \url{https://www.shopify.com/} & A popular e-commerce platform \\
        \bottomrule
    \end{tabular}
\end{table}

To understand how each pair performs, I ran PageSpeed Insights against each site once 5th of November 2025\footnote{PageSpeed results tend to fluctuate over time, especially for content-based websites, since content itself can affect the results and occasionally sites have been improved further by their developers.}.
PageSpeed Insights is a commonly used Google service available at \url{https://pagespeed.web.dev/}.
To gain further insights, I also set up a small Puppeteer\footnote{Puppeteer \url{https://pptr.dev} allows controlling Chrome browser in a headless way and capturing data.} script to gather the amount of JavaScript (JS) and the amount of related requests on each page. 
The runs were performed on the 5th of November 2025, and I have enumerated the mobile results in \autoref{tab:holotype_results}.
Based on the results, the HTML First variants seem to perform better than their comparison points based on this initial measurement.
As might be inferred from the site orientation (HTML or JavaScript), there is a clear difference in how much JavaScript sites request.
In the comparison, HTML-oriented websites barely use JavaScript, which is understandable, while the JavaScript files they use tend to be small in size.
Hacker News stands out as an interesting outlier in terms of accessibility since it has the lowest score in the category, indicating that even in an HTML-oriented approach, developers should follow best accessibility practices, even if the site would otherwise be small and fast.
All of the websites could likely be optimized further in all regards, but in general, there seems to be left work to be done with HTML-oriented websites.


\begin{table}[ht]
    \caption{The table enumerates PageSpeed Insights for each website tested. Each category receives a score from 0 to 100, with 100 considered the best available score within that category. Scores are only indicative, and there can still be, for example, accessibility issues remaining on a site even with a perfect score.}
    \label{tab:holotype_results}
    \begin{tabularx}{\textwidth}{cXssszss}
        \toprule
        Holotype & Site & Perfor\-mance & Accessi\-bility & Best Practices & SEO & Amount of JS (kB) & JS requests \\
        \midrule
        Content & Learn Rust & 100 & 91 & 100 & 100 & 10.2 & 3 \\
        Content & Next.js documentation & 86 & 96 & 96 & 100 & 416.9 & 28 \\
        News & Hacker News & 100 & 52 & 96 & 75 & 73.8 & 5 \\
        News & The Verge & 34 & 93 & 77 & 100 & 2160.6 & 44 \\
        E-commerce & Kirby CMS & 95 & 100 & 100 & 100 & 0 & 2 \\
        E-commerce & Shopify platform & 70 & 88 & 100 & 92 & 346.1 & 15 \\
        \bottomrule
    \end{tabularx}
\end{table}

\subsection{Benchmark against Yle website}

Yle is a frequently visited site, as it is the website of the Finnish broadcasting corporation, and it serves as a good example of a media-oriented website that could benefit from optimization to reduce hosting costs and, you could argue, to provide better public service.
For the Yle site, I followed the steps below to apply HTML First principles:

\begin{enumerate}
    \item I took a snapshot of Yle\footnote{Yle is the Finnish broadcasting corporation. The site is available through \url{https://yle.fi}.} landing page using Chrome browser in incognito mode\footnote{The benefit of saving the page like this is that it avoids any potential code injected by browser plugins I might have enabled on my machine.} to save the page and its related assets so the page can be run locally.
    \item I made a copy of the snapshot to modify.
    \item I adjusted the copy according to HTML First principles. I have described my modifications in more detail below.
    \item I set up Google Lighthouse\footnote{Lighthouse is a popular solution for measuring site performance. See \url{https://developer.chrome.com/docs/lighthouse/} for further information.} to compare the original and modified variants within Chrome browser. In my setup, Lighthouse is run with its default settings against a mobile profile\footnote{In mobile profile differences tend to be more visible than in the desktop one since it is more constrained by definition.}.
\end{enumerate}

The purpose of my modifications was to provide a sense of a potential "best" case for adopting HTML First principles for each site while maintaining sufficient functionality.
The target was to achieve a roughly visually complete port with secondary functionality, such as a cookie banner or videos, missing.
I have enumerated my key changes below to document how the HTML First variant differs from the base one.

\begin{enumerate}
    \item As I noticed that the Yle landing page relies heavily on JavaScript for logic and styling via CSS-in-JS\footnote {CSS-in-JS is a common CSS technique used especially in applications, as it allows using CSS within application code easily.}, I decided to remove JavaScript usage to see how much it would break.
    \item Since removing JavaScript subsequently broke CSS, I wrote a small script to extract CSS as a static file to restore styling. I did minor modifications to drop redundant rules that were overly broad and broke styling\footnote{There's a caveat related to this since CSS-in-JS can model dynamic behavior that might not be possible to replicate trivially through naïve style extraction.}.
    \item I restored the behavior of the main carousel at the top of the page using a small amount of CSS and JavaScript for the previous/next buttons, since it was not possible to implement them with CSS alone due to technical constraints.
\end{enumerate}


It is important to note that the results I received from Lighthouse are only indicative and do not capture the whole picture, as I did not port all the functionality to HTML First style\footnote{It would be a good idea to restore video playing by loading the player code as it is needed, for example.}.
I have enumerated Lighthouse scores in \autoref{tab:variant_comparison}, and the scores show that the modified variant performs consistently better than the original one.
Besides overall scores, Lighthouse provides more detailed performance information, as shown in \autoref{tab:variant_performance}.
Notably, the modified version performs considerably better in each category, likely due to a smaller payload: there is less to transfer, and less work to do since CSS rules do not have to be processed through JavaScript.

\begin{table}[ht]
    \caption{The table compares the variants across Lighthouse scores. Note how the modified variant received consistently better results.}
    \label{tab:variant_comparison}
    \begin{tabular}{cllll}
        \toprule
        Variant & Performance & Accessibility & Best Practices & SEO \\
        \midrule
        Original & 58 & 91 & 57 & 100 \\
        Modified & 90 & 95 & 75 & 100 \\
        \bottomrule
    \end{tabular}
\end{table}

\begin{table}[ht]
    \caption{The table compares the variants based on Lighthouse performance metrics. Lighthouse documentation \cite{chromeLighthousePerformance} explains how to interpret specific values in detail. Still, the modified variant seems to perform better across the board, since there is less work to do due to the reduced JavaScript and computation.}
    \label{tab:variant_performance}
    \begin{tabular}{clllll}
        \toprule
        Variant & FCP (s) & LCP (s) & TBT (ms) & CLS & SI (s) \\
        \midrule
        Original & 4.4 & 10.2 & 310 & 0.01 & 4.4 \\
        Modified & 2.0 & 3.4 & 0 & 0.003 & 2.0 \\
        \bottomrule
    \end{tabular}
\end{table}


\subsection{Summary}

The examples of this section show how focusing on HTML can give performance benefits, especially for content-focused websites, as shown by the holotype-based comparisons and benchmark against the Yle website, but that is not the only benefit.
As shown by the htmx-based examples with SaaS products, moving to a hypermedia-style approach that favors HTML seems to give benefits beyond performance in terms of a smaller codebase that is faster to develop and therefore easier to maintain, as fewer moving parts can break.
HTML First principles call for using the platform, and interestingly, the approach reminds us of the early days of web development before JavaScript-based frameworks became the mainstream approach for developing complex web applications.



\section{Discussion}
\label{sec:discussion}
Although it is hidden to many, HTML is one of the core technologies of the web that, when used correctly, can improve accessibility of websites and applications while even allowing the definition of complex user interfaces.
HTML First development calls for acknowledging the role of HTML and leveraging it to its fullest extent.

\subsection{The friction between web frameworks and the web platform}

As the use of the web expanded to applications since the early 2000s, it has put pressure on the platform and associated standardization processes.
Although the web platform was not initially designed as an application platform, developers repurposed it for that purpose.
Essentially, the web platform had gained sufficient functionality by the early 2000s to become what it is today, and particularly, web frameworks were leading the way by capturing common patterns and easing the development of complex web applications.
That said, eventually, web standards have begun to catch up and capture common use cases within the web platform itself\footnote{This process has not always been without its problems as the slow adoption of Web Components \cite{webcomponents} shows.}.
This process includes HTML, as HTML5 was a culmination of its development, and since then, it has evolved into a living standard with gradual improvements adopted by browsers.
As the platform has caught up, this means web frameworks have less to do, and HTML First development is a part of this.

\subsection{What features should be added to HTML}

As noted by \citet{stateofhtml2024conclusion}, there is a balance between adding new features to HTML and enhancing existing ones.
Standards fix behavior by definition, and given the way web standards work, it can be challenging to remove a feature once it has been added to the standard.
Given that it is challenging to standardize complex behaviors, it may make sense to focus on standardizing and improving core primitives that developers can use to build their components.
W3C's Open UI working group \cite{openui2025} is actively working towards figuring out UI primitives to standardize to the web platform while providing polyfills\footnote{Polyfills are small pieces of code that allow new functionality to work until browsers support it.} for features in progress.
Another way to look at it is to consider especially interactive holotypes \cite{vepsalainen2025potential} as those use cases may inform the design for complex elements that potentially should be added to the HTML standard.
Triptych proposals by \citet{petros2025} give another vision of where HTML could be headed in case it were made more powerful to support hypermedia more fully at a syntax level, as this would no doubt open many new use cases for HTML and likely render libraries like htmx largely obsolete.
In short, it is not easy to add new features to HTML since essentially they cannot be removed, and therefore standardization has to be conservative, although it should aim to add common primitives that extend the usefulness of the HTML standard in considerable ways.



\subsection{HTML First as an antidote to frameworkitis}

Although web frameworks make it easy to create applications, leveraging them does not always mean you are getting the most out of the web platform.
Overuse or misuse of web frameworks could be dubbed as frameworkitis\footnote{Interestingly enough, the term frameworkitis was used by Jeff Atwood as early as 2005 \cite{atwood2005}.}.
It means you are missing out on technical solutions outside of your framework, or may be forced to work against it to go your way.
By emphasizing the capabilities of modern HTML, HTML First development asks developers to consider how far they can go with HTML, CSS, and, finally, JavaScript before considering other options, to stay close to the platform.
Adoption of HTML First thinking does not mean you should abandon your framework, but rather it calls for being aware of what the web platform can do for you.
The HTML First approach also asks you to consider your dependencies and whether you could do more with less.

\subsection{HTML First encourages locality of behavior}

HTML First encourages developers to think locally, i.e., at the level of a button, allowing them to find related code easily without having to navigate multiple files.
This locality of behavior eases navigating a codebase, although it does not disallow abstraction in the form of Web Components \cite{webcomponents}, for example.
Web Components allow the creation of custom elements to extend HTML semantics \cite{webcomponents} and may be a good way to move beyond basic primitives provided by HTML.

\subsection{Potential maintenance benefits of HTML First}

Since locality of behavior encourages developers to couple related functionality, it eases maintenance, especially in the form of discoverability and understandability of code, as there is less context to consider.
This benefit adds up over time as web applications are often developed over an extended period of time.
In short, HTML First challenges prior notions of what should be units in web development and what should be coupled and what should not.

Although web frameworks may accelerate the development of complex web applications, especially early on, they come with maintenance costs as web technologies evolve.
It can be argued that, given the HTML First approach emphasizes leveraging the platform, maintenance costs are also decreased since there is less to maintain due to framework or library changes\footnote{This is a particularly concrete risk given that the JavaScript ecosystem tends to evolve at a rapid pace and even half a year is a long time. Therefore, developers have to keep track of their dependencies and take care to keep their projects up to date to ensure they have the latest fixes, which occasionally may be security related, especially for backend and full-stack frameworks that span both client and server.}.
HTML First focus reduces the distance between what is considered native and what is considered abstract, thereby leading to fewer moving parts to maintain.

As shown by the examples in \autoref{sec:comparison}, there can be clear benefits in focusing on HTML as this focus may reduce the size and complexity of your website while making it more maintainable and easier to develop features.
Once a feature reaches web standards, it generally stays there, and even websites created in the early days work well today.
Therefore, developing against the web platform and leveraging basic technologies, even within the context of web frameworks, can be a good idea since these technologies are unlikely to break or disappear from web browsers due to how web standardization works. 

\subsection{HTML First is orthogonal to artificial intelligence-driven web development}

As shown by \citet{pohjalainen2025artificial}, Artificial Intelligence (AI) is beginning to impact web development practices.
The way I see it, the HTML First approach is orthogonal to AI-driven development and can be considered separately from it.
However, AI-driven development can likely be combined with HTML First practices for good effect.
That said, it is a good question whether AI could be used to encourage HTML First practices and how to achieve this.
First, good examples of HTML First code should likely be catalogued as a training set to generalize from, and perhaps this kind of limited, specialized model could then be used to spot opportunities for using HTML First practices.
It is an even bigger question whether AI could be used to suggest refactorings that follow HTML First principles, and how this could work out.

\subsection{Adopting HTML First may improve site performance}

As shown by the comparisons in \autoref{sec:comparison}, adopting HTML First principles may yield benefits, such as improved site performance, by encouraging adoption of lightweight approaches, such as favoring CSS over CSS-in-JS.
The main point is that HTML First principles remind us what is essential about websites and encourage us to keep our technical decisions close to the platform, unlocking secondary benefits, like better performance.
However, HTML First alone is not enough for the best outcomes\footnote{In practice, you have to consider other parts of your stack as well to attain good performance and apply specific optimization techniques, such as caching \cite{vepsalainen2023overview}.}.

\subsection{Importance of HTML for AI agents and accessibility}

Clean, well-defined HTML has particular value for the perception of AI agents operating on webpages, as pointed out by \citet{ning2025survey}.
Although accessibility is an increasingly important aspect of websites due to legal movements, such as the EU's Accessibility Act \cite{marcus2022eu}, the growing use of AI agents further emphasizes the need to produce accessible HTML.
Emphasizing the importance of HTML through practices, like HTML First development, is therefore in line with these current trends and highlights the fact that the quality of HTML should be kept in mind during development.

\subsection{Threats to validity}

There are both internal and external threats to validity for the comparisons presented in this paper.
The case studies selected from the htmx project may come with bias as they miss occasions where htmx was not a good fit, and therefore their observations may not be generalizable, or they may come with hidden costs that were not well documented.
For the pairwise comparisons, there may be some selection bias in place, given that there are likely may be other possible pairings and therefore the results should be considered only indicative.
For the Yle case, it is possible the comparison is not exactly fair since some functionality is missing, so the results should be considered only indicative, as they primarily highlight the cost of using a css-in-js solution for styling.
Likely, the observed benefits of the HTML First approach may not be visible in all contexts, such as logic-heavy holotypes including immersive applications, since there is less to optimize by definition, and they tend to have plenty of client-side code by definition.

\section{Conclusion}
\label{sec:conclusion}
HTML First development is a relatively new web development approach described by \citet{htmlfirst2023} in 2023.
As you might tell from the name, the approach emphasizes focusing on HTML before considering other options.
In this article, I explored the topic while researching answers to two research questions I will revisit next, before discussing what could be done next.

\subsection{RQ1: What were the factors leading to the underuse of HTML features in web application development?}

It is practically impossible to develop websites or web applications without even a modest amount of HTML.
The question is more about how HTML is used, and with the development of web applications since the early 2000s, powerful abstractions have become available for web developers.
With these abstractions, it became easy to forget the underlying web platform.
Essentially, web frameworks addressed the deficiencies in the web platform, which was not initially designed for application development, and, understandably, they came with scaffolding that helped developers with development efforts.
Over time, the web standards caught up with this development to an extent and integrated commonly used functionality into the platform itself.
It was this mismatch, and perhaps lack of knowledge of the platform's new features, that led to the underuse of HTML features for web application development.

\subsection{RQ2: How can HTML First development remedy the underuse of HTML, and why is it useful?}

HTML First development emphasizes literally using HTML first before considering other options.
The approach comes with other guiding principles as well, and it strives towards minimalism in development to harken back to the early days of web development when technology was simple and understandable.
HTML First development calls for reducing the amount of abstraction while acknowledging that HTML can achieve more than most developers realize.
There is early evidence that focusing on HTML can give a multitude of benefits for projects, especially in terms of maintenance, as focusing on HTML and complementary approaches, such as leveraging hypermedia, can reduce the amount of code in a project while making it simpler.
This simplicity is valuable since having fewer dependencies generally means developers have less work to do in keeping their project up to date and fewer concepts to consider when developing new features.
As a side effect, adopting HTML First on an existing project may improve performance, especially if older and more inefficient development approaches are replaced with ones closer to the platform.

\subsection{Open research questions}

In this article, we gained a view on what HTML First development is and how it can be valuable for web developers.
Although we already understand something about the topic, addressing several questions could give further insight into the topic and may be worth pursuing as further work:

\begin{enumerate}
    \item What are specific common development patterns (i.e., components, element composition, hypermedia) that capture the ethos of HTML First development well?
    \item What is the magnitude of savings in terms of maintenance costs when comparing HTML-focused projects to JavaScript-heavy ones including more technology?
    \item How easy is it to add new features to an HTML First-inspired codebase vs. a JavaScript-heavy one in terms of effort?
    \item How feasible is it to train AI to perform HTML First-inspired refactorings?
    \item How well do developers know the latest HTML features?\footnote{Yearly State of HTML survey (\url{https://stateofhtml.com/}) gives some idea already while capturing a longitudinal view on the topic.}
    \item What are the main features missing from HTML to make it more useful for web application development?
\end{enumerate}

\subsection{Acknowledgments}

Thanks to Tony Ennis, Carl Kubglenu, Loren Stewart, and Petri Vuorimaa for useful feedback that helped to shape the paper.

\subsection{AI disclaimer}

I used ChatGPT to ideate pairings for the holotype benchmark, as it was able to identify suitable sites effectively, and I validated that each pair made sense within its holotype.4
I also used ChatGPT to generate a small script to figure out the number of JavaScript requests and the size per website, and I used the tool to write a script to extract the styling of a webpage as a standalone CSS file.
I used the Grammarly service for improving the grammar of this paper, and I take responsibility for any grammar mistakes that might remain.



\bibliographystyle{unsrtnat}
\bibliography{references/web}


\section*{Biography}

\medskip
\noindent
\fbox{
\includegraphics[width=3cm]{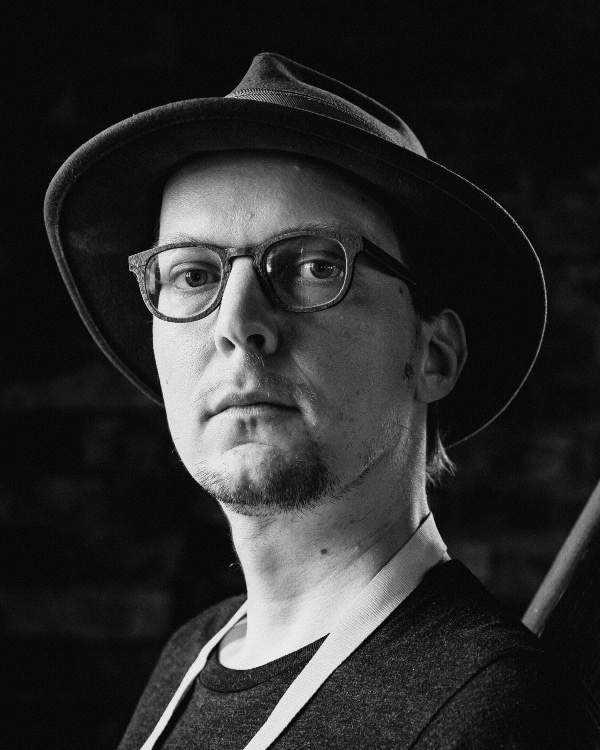}
}

\medskip
\noindent
{\bf Juho Vepsäläinen} (DSc) is a University Teacher at Aalto University, Finland. His current research interests include web development, web performance, and green computing.

\end{document}